\documentclass[apj]{emulateapj}

\usepackage{graphicx,times}
\usepackage{subfigure}
\newcommand{\be}{\begin{equation}}
\usepackage{threeparttable}
\usepackage{booktabs}
\newcommand{\ee}{\end{equation}}
\newcommand{\bea}{\begin{eqnarray}}
\newcommand{\eea}{\end{eqnarray}}

\usepackage{amsmath}
\usepackage{cases}
\usepackage{longtable}
\usepackage{hyperref}
\usepackage{epstopdf}
\usepackage{amsmath,bm}
\usepackage{amssymb}
\usepackage{natbib}
\usepackage{morefloats}
\usepackage{multirow}
\usepackage{array}
\usepackage{verbatim}

\begin{document}

\title{Interpretation of the structure function of rotation measure in the interstellar medium}

\author{Siyao Xu\altaffilmark{1} and Bing Zhang \altaffilmark{1,2,3}}

\altaffiltext{1}{Department of Astronomy, School of Physics, Peking University, Beijing 100871, China; syxu@pku.edu.cn}
\altaffiltext{2}{Kavli Institute for Astronomy and Astrophysics, Peking University, Beijing 100871, China}
\altaffiltext{3}{Department of Physics and Astronomy, University of Nevada Las Vegas, NV 89154, USA; zhang@physics.unlv.edu}

\begin{abstract}

The observed structure function (SF) of rotation measure (RM) varies as a broken power-law function of angular scales.
The systematic shallowness of its spectral slope is inconsistent with the standard Kolmogorov scaling. 
This motivates us to examine the statistical analysis on RM fluctuations. 
The correlations of RM constructed by
\citet{LP15}
are demonstrated to be adequate in explaining the observed features of RM SFs 
through a direct comparison between the theoretically obtained and observationally measured SF results. 
By segregating the density and magnetic field fluctuations and adopting arbitrary indices for their respective power spectra, 
we find that when the SFs of RM and emission measure have a similar form over the same range of angular scales, 
the statistics of the RM fluctuations reflect the properties of density fluctuations.
RM SFs can be used to evaluate the mean magnetic field along the line of sight, 
but cannot serve as an informative source on the properties of turbulent magnetic field in the interstellar medium.
We identify the spectral break of RM SFs as the inner scale of a shallow spectrum of electron density fluctuations, which 
characterizes the typical size of discrete electron density structures in the observed region. 
\end{abstract}

\keywords{ISM: magnetic fields -- turbulence -- methods: statistical}

\section{Introduction}

Astrophysical turbulence, in spite of the stochastic nature, 
allows for a statistical study that can have access to its underlying regularities
\citep{Biskampbook, Brad13, BL15}.
The turbulent spectrum, as a statistical measure of turbulence, contains a great wealth of information on the injection, nonlinear transfer, and 
dissipation of turbulent energy, and thus can characterize the essential properties of interstellar turbulence.
The statistics of the turbulent velocity provide a direct diagnostic of the turbulent spectrum, but it is challenging to disentangle velocity and 
density contributions when utilizing spectroscopic data to obtain velocity statistics
\citep{Ves03}.
Among the attempts to overcome this difficulty, 
new techniques, e.g., the Velocity Channel Analysis and Velocity Coordinate Spectrum, have been developed on a solid theoretical ground 
and successfully tested by numerical simulations
(see a review by \citealt{Laz09rev}).
On the other hand, the statistical study of density is rather straightforward and has attracted more attention
\citep{GN85, Spa90, Armstrong95, CL10}. 
A Kolmogorov spectrum of the fluctuations in the interstellar plasma density is suggested by observational evidence. 
However, density is a passive scalar and the measure of density fluctuations 
can only be regarded as an indirect approach of tracing turbulence. 
Numerical studies show that 
the density spectrum significantly deviates from the velocity spectrum in supersonic turbulence 
\citep{CL03, BLC05,KL07}.

Turbulence induces fluctuating magnetic field by the small-scale dynamo, and brings the magnetic energy up to 
the injection scale of turbulence through the inverse cascade
\citep{CVB09, Bere11, Zra14}.
The generated magnetic field in turn affects the properties of turbulence and converts the hydrodynamic turbulence 
into magnetohydrodynamic (MHD) turbulence, which is a common state of interstellar plasma. 
Achieving the spectral profile of turbulent magnetic field is crucial for 
studying the processes such as cosmic-ray scattering
\citep{Kota_Jok2000},
star formation
\citep{Mck99},
and magnetic reconnection 
\citep{LV99}.
It requires an adequate understanding of MHD turbulence. 
The point of contention is whether the scaling law for hydrodynamic turbulence is still valid 
in the context of MHD turbulence. 
Within theory's reach, 
\citet{GS95} 
pointed out that the transverse mixing motions of magnetic field lines in MHD turbulence preserve the character of 
hydrodynamic turbulent motions, 
thus the hydrodynamic turbulence scaling holds in the direction perpendicular to the local magnetic field. 
The Kolmogorov-type spectrum of density fluctuations in the magnetized interstellar plasma is in accordance with this theoretical expectation. 
Statistical analyses of the magnetic field data produced by numerical simulations support the theory
\citep{CV00, CLV_incomp,MG01}. 
However, there is a shortage of observational evidence since
magnetic field statistics are more poorly constrained from observations than velocity and density 
statistics. 
Magnetic field cannot be measured independently, but is intermixed with other quantities such as densities of relativistic 
electrons or thermal electrons. 
Only a theoretical model capable of reproducing the detected features of related observables 
can give us confidence in eliminating the inherent ambiguities and unveiling the physics in the measurements of turbulent 
magnetic fields.

Based on the modern understanding of MHD turbulence, 
\citet{LP12, LP15}
carried out comprehensive statistical studies
on fluctuations in synchrotron intensity, synchrotron polarization, and Faraday measure. 
They provided a thorough exposition on the quantitative correlations between the 
statistics of synchrotron emission and characteristics of the underlying magnetic turbulence. 
Their synchrotron studies of turbulence and cosmic magnetic fields 
open the avenue to a wide range of astrophysical applications. 
In particular, 
\citet{LP15} (hereafter LP16)
presented the structure function (SF) analysis of rotation measures (RMs), including both cases with 
spatially-coincident and spatially-separated synchrotron emission and Faraday rotation regions. 
The latter case can be applied to probing the turbulent magnetic fields 
embedded in the diffuse ionized component of the interstellar medium (ISM), when the observed Faraday rotation 
only contains the contribution from the Galaxy.

In practice, the SFs of RMs have been attained from a number of independent observations covering 
different scales and areas in the Galaxy
(e.g., \citealt{Sim84,Sim86,Cle92,MS96,Hav03,Hav04,SH04,Hav08,Roy08, Stil09, Opp12};
see also, \citealt{Han04,Han09}).
In combination with the SFs of emission measures (EMs), it provides a possibility for determining the properties of turbulent magnetic fields in the ISM. 
The observations reveal some common features in the form of RM SFs as a function of angular scales:
(1) The SF has a much shallower slope than that expected from the standard Kolmogorov power law. 
(2) The SF follows a broken power spectrum changing from a relatively steeper slope to a shallower one at a scale 
on the order of $1$ pc. 
(3) The slope of the SF varies from region to region and has a dependence on Galactic latitude.
(4) The slope of the SF tends to flatten at large angular scales. 
(5) The SF of EMs shows a similar slope to that of RMs detected from the same region.
There has not been a compelling interpretation for these features in earlier literature. 
As an empirical attempt, 
\citet{MS96} 
suggested
that the broken power-law spectrum may result from the transition from three-dimensional to two-dimensional filamentary turbulent 
structure, but this imposed turbulent structure, which may not be a common occurrence, 
fails to account for other observational features.

As the major impediment of the problem, 
both density and magnetic field fluctuations are imprinted in the observed RM fluctuations. 
The relative importance between them determines 
whether the behavior of RM SFs can effectively diagnose the turbulent magnetic fields. 
In the studies of, e.g. 
\citet{Sim84, LP15}, 
the product of electron density and magnetic field has been treated as a composite quantity. 
In 
\citet{MS96}, 
the two are separated, but both fluctuations are assumed to conform to the Kolmogorov spectrum. 
In the current work, in order to resolve the respective influence of density and magnetic field fluctuations 
and gain a clear insight into the properties of their associated turbulence, 
we separate the density and magnetic field components in the statistical analysis of RM fluctuations, 
but do not restrict the scalings of their respective power spectra. 
Our goal is not only to seek understanding of the observed features of RM SFs, 
but also to clarify its relation with the turbulent density field and magnetic field.
In Section 2, we present a statistical analysis of RM and EM fluctuations, and provide the expressions of 
their SFs with separate contributions from electron density and magnetic field.
In Section 3, we compare the analytical result with the measured RM and EM SFs from observations. 
Conclusions and discussions are given in Section 4.

\section{Statistical analysis of RM and EM fluctuations}

As the fundamental radio propagation measurements, the measure of magnetization (RM) and electron densities 
(EM) provide unique information on the magnetized turbulence in the diffusive, ionized component of the ISM. 
In astronomically convenient units, they are defined as 
\begin{equation}
  \text{RM} (\text{rad m}^{-2})= 0.81 \int_0^L n_e (\text{cm}^{-3}) B_z ( \mu \text{G}) dz (\text{pc}),
\end{equation}
and
\begin{equation}
  \text{EM} (\text{pc cm}^{-6}) = \int_0^L  [n_e (\text{cm}^{-3})]^2 dz (\text{pc}),
\end{equation}
where $n_e$ is electron density, $B_z$ is the line-of-sight (LOS) component of magnetic field, and $L$ is the path length 
through the Faraday rotating medium.

\subsection{SFs of RM and EM fluctuations}\label{sec: lp}

For our statistical analysis of RM and EM fluctuations, we follow the approach employed by 
LP16
that deals with
RM per unit length along LOS, namely, RM density $0.81(n_eB_z)$, and EM density $(n_e^2)$. 
We first treat them as composite quantities. 
We consider them to be statistically homogeneous and isotropic. 
This ensures that these quantities are invariant with respect to the LOS orientation. 

We assume that the RM (EM) density can be described as a sum of its ensemble-average mean and zero mean fluctuations, 
\begin{equation}\label{eq: rmdendes}
    \phi(\bm{X}, z) = \phi_0 + \delta \phi (\bm{X}, z),   ~~ \langle \delta \phi(\bm{X}, z) \rangle =0, 
\end{equation}
where $\bm{X}$ denotes the position on the plane of sky and $z$ is the distance along the LOS. 
We use $\langle ... \rangle$ to denote an ensemble average. 
As the real-space statistical tool, the two-point correlation function (CF) is  
\begin{equation}
    \xi(\bm{R}, \Delta z) = \langle  \phi(\bm{X_1}, z_1) \phi(\bm{X_2}, z_2) \rangle,  
\end{equation}
for RM (EM) density, and 
\begin{equation}
  \widetilde{\xi}(\bm{R}, \Delta z) =  \langle  \delta \phi(\bm{X_1}, z_1) \delta \phi(\bm{X_2}, z_2) \rangle
\end{equation}
for RM (EM) density fluctuations. 
The two are related as 
\begin{equation}
  \widetilde{\xi}(\bm{R}, \Delta z) =  \xi(\bm{R}, \Delta z) -  \phi_0^2, 
\end{equation}
with $\bm{R} = \bm{X_1}-\bm{X_2}$ and $\Delta z = z_1-z_2$. 
The SF of RM (EM) density and RM (EM) density fluctuations are identical, 
\begin{equation}\label{eq: rmdsf}
       d(\bm{R}, \Delta z) = \widetilde{d}(\bm{R}, \Delta z) = 
     \langle  [\delta \phi(\bm{X_1}, z_1)- \delta \phi(\bm{X_2}, z_2)]^2    \rangle.      
\end{equation}

According to the statistical descriptions presented in
LP16, 
we adopt a power-law model of CF and SF, which is adequate for characterizing the scaling properties of turbulence. 
Their forms are 
\begin{subequations}\label{eq: cfsflp}
\begin{align}
   & \widetilde{\xi}_\phi (\bm{R},\Delta z) = \sigma_\phi^2 \frac{r_\phi^{m_\phi}}{r_\phi^{m_\phi} + (R^2 + \Delta z^2)^{m_\phi/2}}, \label{eq: molcf}\\
   & \widetilde{d}_\phi (\bm{R},\Delta z) = 2  \sigma_\phi^2  \frac{(R^2 + \Delta z^2)^{m_\phi/2}}{r_\phi^{m_\phi} + (R^2 + \Delta z^2)^{m_\phi/2}}, 
\end{align}
\end{subequations}
with the variance of fluctuations defined as 
\begin{equation}
   \sigma_\phi^2 = 0.81^2 \langle \delta(n_e B_z)^2   \rangle
\end{equation}
for RM density, and 
\begin{equation}
    \sigma_\phi^2 = \langle \delta(n_e^2)^2 \rangle
\end{equation}
for EM density, 
where $r_\phi$ is the correlation scale of RM (EM) density fluctuations, and 
$m_\phi$ is the index of their power-law functions in real space.
Under the condition of statistical homogeneity, 
we see from above expressions that CF and SF only depend on the relative separation distance 
instead of the separation vector between the two points.

The power spectrum in Fourier space $E(k) \sim k ^\alpha$ is complementary to CF and SF. 
Since CF and SF respectively apply to small-scale and large-scale dominated statistics, 
a shallow ($\alpha>-3$) spectrum is more properly described by CF, while a steep ($\alpha <-3$) spectrum more favors SF treatment.
Only when the cutoffs of CF at small scales and SF at large scales are both defined, can CF and SF be related and employed simultaneously
(see detailed discussions in \citealt{LP04,LP06}).
From Eq. \eqref{eq: cfsflp}, we find 
\begin{equation}
\begin{aligned}
&    \widetilde{\xi}_\phi (0) = \sigma_\phi^2, \\
&    \widetilde{d}_\phi (\infty) =  2  \sigma_\phi^2,
\end{aligned}
\end{equation}
and thus the CF and SF are related by
\begin{equation}
\begin{aligned}
   & \widetilde{d}_\phi (\bm{R},\Delta z) = 2[\widetilde{\xi}_\phi (0) - \widetilde{\xi}_\phi (\bm{R},\Delta z)], \\
   & \widetilde{\xi}_\phi (\bm{R},\Delta z) = \frac{1}{2} [\widetilde{d}_\phi (\infty) - \widetilde{d}_\phi (\bm{R},\Delta z)].
\end{aligned}
\end{equation}
The relation between the CF (SF) index $m_\phi$ and the spectral index $\alpha$ depends on whether the turbulent spectrum is shallow or steep 
\citep{LP06}, 
\begin{subnumcases}
{\alpha = \label{eq: malp}}
  m_\phi-N, ~~~~~~ \alpha >-3, \label{eq: alaslw} \\
  -m_\phi-N, ~~~\alpha <-3, 
\end{subnumcases}
where $N$ is the dimensionality of space. 
In the case of three-dimensional Kolmogorov turbulence with $\alpha = -11/3$, the corresponding value of $m_\phi$ is $2/3$. 
Regarding the correlation scale $r_\phi$, as pointed out by LP16, 
it corresponds to the energy dissipation scale for a shallow spectrum at wavenumbers smaller than $1/r_\phi$, and 
the injection scale of turbulent energy for a steep spectrum at wavenumbers larger than $1/r_\phi$. 
Strictly speaking, the forms of CF and SF given in Eq. \eqref{eq: cfsflp} are only applicable in the inertial range of the spectrum. 
For scales below the inner scale $r_\phi$ of a shallow spectrum and above the outer scale $r_\phi$ of a steep spectrum, 
the exact forms of CF and SF depend on the specific dissipation and injection processes of turbulent energy.

The total RM and EM are the integrals of RM and EM densities over the path length along the LOS
and thus their SFs have a dependence on the integration path. 
In a simple case where there is only one single Faraday rotating screen along the LOS with a thickness $L$, as derived in 
LP16, 
the SF for the fluctuations of RM (EM) is given by  
\begin{equation}\label{eq: sflp}
    D_{\Delta{\Phi}}(\bm{R},L,L) = 2 \int_0^L  d \Delta z (L-\Delta z) [\widetilde{\xi}_\phi(0,\Delta z)-\widetilde{\xi}_\phi (\bm{R},\Delta z)]. 
\end{equation}
Notice that the SF for RM (EM) has the same form as above due to 
$\widetilde{\xi}_\phi(0,\Delta z)-\widetilde{\xi}_\phi (\bm{R},\Delta z) =\xi_\phi(0,\Delta z)-\xi_\phi (\bm{R},\Delta z)$. 
A more complicated expression of $D_{\Delta{\Phi}}$ applicable to the situation with the synchrotron radiation and 
Faraday rotation taking place in the same volume is available in 
LP16.

We consider a thick Faraday screen with $L>r_\phi$, which is common to extragalactic sources.
After inserting Eq. \eqref{eq: molcf}, 
the SF from Eq. \eqref{eq: sflp} has asymptotic expressions in different ranges of $R$ 
(see Appendix C in LP16),
\begin{subnumcases}
     { D_{\Delta{\Phi}}(R) = \label{eq: appc}}
       2 \sigma_\phi^2 LR \Big(\frac{R}{r_\phi}\Big)^{m_\phi}, ~~~~~~R<r_\phi<L,\\
       2 \sigma_\phi^2 LR \Big(\frac{R}{r_\phi}\Big)^{-m_\phi}, ~~~~ r_\phi<R<L,\\
       2 \sigma_\phi^2 L^2 \Big(\frac{L}{r_\phi}\Big)^{-m_\phi}, ~~~~~ r_\phi<L<R.
\end{subnumcases}
Regarding its dependence on $R$, the slope changes from $1+m_\phi$ to $1-m_\phi$ when $R$ reaches the correlation scale $r_\phi$, 
and flattens when $R$ exceeds $L$.  
It is necessary to point out that, as we discussed above, 
the expression of $D_{\Delta{\Phi}}(R)$ at $R<r_\phi$ in the case of a shallow spectrum of RM (EM) density fluctuations 
and those at $R>r_\phi$ in the case of a steep spectrum are not robust. 
In the following calculations 
we assume that the same power-law model of $\widetilde{\xi}_\phi (\bm{R},\Delta z)$ can be extensively applied beyond the inertial range of turbulence, 
keeping in mind that $m_\phi$ has an adjustable value in different ranges of scales, 
and we will discuss the modifications in a more realistic situation in Section \ref{sec: comobs}.

For an observer sitting in the Galaxy, if the Faraday rotation effect for the extragalactic sources mainly arises from the Galaxy, 
the angular separation $\theta$ between a pair of LOSs through the ISM 
coincides with the ratio of the projected distance and 
geometrical depth of the Galactic Faraday material, i.e., $R/L$. 
Thus, Eq. \eqref{eq: appc} can be recast into
\begin{subnumcases}
     { D_{\Delta{\Phi}}(\theta) = \label{eq: comsfthe}}
       2 \sigma_\phi^2 L^2 \Big(\frac{L}{r_\phi}\Big)^{m_\phi}\theta^{1+m_\phi}, \label{eq: comrmsa}   \nonumber \\
       ~~~~~~~~~~~~~~~~~~~~~~~~~~~~~~~~~~~\theta<\frac{r_\phi}{L}<1,  \\
       2 \sigma_\phi^2 L^2 \Big(\frac{L}{r_\phi}\Big)^{-m_\phi} \theta^{1-m_\phi}, \label{eq: comrmla}  \nonumber \\
      ~~~~~~~~~~~~~~~~~~~~~~~~~~~~~~~~~~ \frac{r_\phi}{L}<\theta<1,   \\
       2 \sigma_\phi^2 L^2 \Big(\frac{L}{r_\phi}\Big)^{-m_\phi}, \label{eq: comrmlla} \nonumber \\
       ~~~~~~~~~~~~~~~~~~~~~~~~~~~~~~~~~~ \theta>1>\frac{r_\phi}{L}. 
\end{subnumcases}
If the underlying turbulence conforms to the Kolmogorov scaling, there is $m_\phi = 2/3$, so we expect that the SF exhibits a 
broken slope changing from $5/3$ at $\theta < r_\phi/L$ to $1/3$ at $r_\phi/L < \theta <1$, 
and remains unchanged at $\theta>1$.

The above conversion from the linear scale $R$ to angular scale $\theta$ should be adjusted
when the LOSs intersect a Faraday screen at a distance from the observer.
The observed angular separation of a source pair becomes
\begin{equation}
  \theta = \frac{R}{L+L_f},
\end{equation}
where $L_f$ is the distance from the Faraday screen to the observer. 
By substituting
\begin{equation}
   L^\prime = L + L_f,  ~~\zeta =\frac{L}{L^\prime},
\end{equation}
and inserting 
\begin{equation}
   R = L^\prime \theta,
\end{equation}
Eq. \eqref{eq: appc} is reformulated as 
\begin{subnumcases}
     { D_{\Delta{\Phi}}(\theta) = \label{eq: extrintr}}
       2 \sigma_\phi^2 \zeta {L^\prime}^2 \Big(\frac{L^\prime}{r_\phi}\Big)^{m_\phi}\theta^{1+m_\phi},    \nonumber \\
       ~~~~~~~~~~~~~~~~~~~~~~~~~~~~~~~~~~~\theta<\frac{r_\phi}{L^\prime}<\zeta,  \\
       2 \sigma_\phi^2 \zeta {L^\prime}^2 \Big(\frac{L^\prime}{r_\phi}\Big)^{-m_\phi} \theta^{1-m_\phi},   \nonumber \\
      ~~~~~~~~~~~~~~~~~~~~~~~~~~~~~~~~~~ \frac{r_\phi}{L^\prime}<\theta<\zeta,   \\
       2 \sigma_\phi^2 L^2 \Big(\frac{L}{r_\phi}\Big)^{-m_\phi},  \nonumber \\
       ~~~~~~~~~~~~~~~~~~~~~~~~~~~~~~~~~~ \theta>\zeta>\frac{r_\phi}{L^\prime}. 
\end{subnumcases}
The actual value of $\zeta$ can be evaluated from observations as the angular scale beyond which 
$D_{\Delta{\Phi}}(\theta)$ has a zero slope. 
We notice that Eq. \eqref{eq: comsfthe} corresponds to the specialization of the above equation at $\zeta =1$, 
applicable to both extragalactic sources with little internal Faraday rotation and Galactic sources.
In the opposite limit, for an extended extragalactic source with high internal Faraday rotation 
(within which the multiple RM components are correlated)
and relatively negligible Galactic contribution, 
$\zeta$ can be much less than unity, 
from which and the estimated size of the radiation emitting region ($\sim L$), 
the location of the Faraday screen, i.e., the distance of the source, can potentially be obtained.
But such an observation requires very high angular resolution due to the large distance of the extragalactic source.

Besides the case with a single thick Faraday screen, which is the focus of this paper,
analyses for other realizations with, e.g., a thin Faraday screen ($L<r_\phi$) or a single LOS ($R=0$) are also 
provided in 
LP16,
which can be widely applied to different observational situations.

\subsection{SFs with separate contributions from electron density and magnetic field} \label{sec: thw}

The above approach straightforwardly reveals the dependence of the 
slope and amplitude of the RM SF on the spectral characteristics of RM density fluctuations. 
But it has the disadvantage that from the observed RM SF, 
the relative significance between density and magnetic field fluctuations cannot be readily discerned.
We next carry out an analogous derivation of SFs of RM and EM as above, but 
separate the contributions from density and magnetic field. 

Similarly, we assume that the electron density and LOS component of magnetic field are described by 
\begin{equation}
  n_e= n_{e0} + \delta n_e, ~~ B_z = B_{z0} + \delta B_z,
\end{equation}
and 
\begin{equation}
  \langle \delta n_e \rangle =0, ~~ \langle \delta B_z \rangle =0, 
\end{equation}
such that we have the product $n_e B_z$ 
\begin{equation}
\begin{aligned}
    n_e B_z &= (n_{e0} + \delta n_e ) (B_{z0} + \delta B_z ) \\
                                      &= n_{e0} B_{z0} + n_{e0} \delta B_z + B_{z0} \delta n_e + \delta n_e  \delta B_z, 
\end{aligned}                                     
\end{equation}
and the squared $n_e$ 
\begin{equation}
   n_e^2 = n_{e0}^2 + \delta n_e  ^2 + 2n_{e0} \delta n_e .  
\end{equation}
It follows that the CFs of RM and EM densities become (see detailed calculations given in the Appendix)
\begin{equation}\label{eq: seprm}
  \xi_\phi(\text{RM}) = 0.81^2 (n_{e0}^2 B_{z0}^2 + n_{e0}^2  \widetilde{\xi}_B +  B_{z0}^2  \widetilde{\xi}_n +  \widetilde{\xi}_{n}  \widetilde{\xi}_B), 
\end{equation}
and 
\begin{equation}\label{eq: sepem}
  \xi_\phi(\text{EM})  = (n_{e0}^2 + \langle \delta n_e^2   \rangle)^2 + 2  \widetilde{\xi}_n^2 + 4n_{e0}^2  \widetilde{\xi}_n .
\end{equation}
Here $ \widetilde{\xi}_n$ and $ \widetilde{\xi}_B$ are CFs for fluctuations in $n_e$ and $B_z$. 
By adopting the same model of CF as introduced in Eq. \eqref{eq: molcf}, we have 
\begin{equation}
\begin{aligned}
      \widetilde{\xi}_n (\bm{R},\Delta z) &= \langle \delta n_e (\bm{X_1}, z_1)\delta n_e (\bm{X_2}, z_2) \rangle \\
  & =\sigma_n^2 \frac{r_n^{m_n}}{r_n^{m_n} + (R^2 + \Delta z^2)^{m_n/2}}, \\
      \widetilde{\xi}_B (\bm{R},\Delta z) &=\langle \delta B_z (\bm{X_1}, z_1)\delta B_z (\bm{X_2}, z_2) \rangle \\
  & =\sigma_B^2 \frac{r_B^{m_B}}{r_B^{m_B} + (R^2 + \Delta z^2)^{m_B/2}},
\end{aligned}
\end{equation}
with their respective correlation lengths $r_n, r_B$, power-law indices 
$m_n, m_B$, and variances of fluctuations 
\begin{equation}
     \sigma_n^2  = \langle \delta n_e^2   \rangle, ~~~
     \sigma_B^2  = \langle \delta B_z^2 \rangle. 
\end{equation}

Under the condition of relatively small fluctuations, the linear terms in $\xi_\phi(\text{RM})$ (Eq. \eqref{eq: seprm}) and 
$\xi_\phi(\text{EM})$ (Eq. \eqref{eq: sepem}) play a dominant role in determining the resultant SFs of RM and EM. 
Combining Eq. \eqref{eq: sflp} with Eq. \eqref{eq: seprm} and \eqref{eq: sepem}, we obtain analytical estimates, 
\begin{equation}\label{eq: drmanaes}
\begin{aligned}
    & D_\text{RM} \approx \\
      & 2 \times 0.81^2 n_{e0}^2 \int_0^L  d \Delta z (L-\Delta z) [\widetilde{\xi}_B(0,\Delta z)-\widetilde{\xi}_B (\bm{R},\Delta z)]  \\
      & + 2 \times 0.81^2 B_{z0}^2 \int_0^L  d \Delta z (L-\Delta z) [\widetilde{\xi}_n(0,\Delta z)-\widetilde{\xi}_n (\bm{R},\Delta z)] ,
\end{aligned}
\end{equation}
and 
\begin{equation}\label{eq: sepsfem}
 D_\text{EM} \approx  2 \times  4n_{e0}^2 \int_0^L d\Delta z (L-\Delta z) \Big[ \widetilde{\xi}_n(0,\Delta z) - \widetilde{\xi}_n(\bm{R},\Delta z)\Big]   .
\end{equation}
By comparing Eq. \eqref{eq: sflp}, \eqref{eq: comsfthe}, and \eqref{eq: drmanaes}, 
the simplified expressions of $D_\text{RM}$ in different asymptotic regimes can then be derived:
\begin{subnumcases}
     { D_\text{RM} (\theta)= \label{eq: msclex}}
       2 \times 0.81^2 L^2 \Big [n_{e0}^2 \sigma_B^2  \Big(\frac{L}{r_B}\Big)^{m_B}\theta^{1+m_B}  \nonumber \\
       ~~~~~~~~~~~~~~~~~+ B_{z0}^2 \sigma_n^2  \Big(\frac{L}{r_n}\Big)^{m_n}\theta^{1+m_n} \Big ], \nonumber \\
        ~~~~~~~~~~~~~~~~~~~~~~~~~~~~\theta<\frac{r_n}{L}<\frac{r_B}{L}<1,    \\
       2 \times 0.81^2 L^2 \Big [n_{e0}^2 \sigma_B^2  \Big(\frac{L}{r_B}\Big)^{m_B}\theta^{1+m_B}  \nonumber \\
       ~~~~~~~~~~~~~~~~~+ B_{z0}^2 \sigma_n^2 \Big(\frac{L}{r_n}\Big)^{-m_n} \theta^{1-m_n} \Big],   \nonumber \\
        ~~~~~~~~~~~~~~~~~~~~~~~~~~~~\frac{r_n}{L}<\theta<\frac{r_B}{L}<1,   \label{eq: sepdrmb} \\
       2 \times 0.81^2 L^2 \Big [n_{e0}^2 \sigma_B^2 \Big(\frac{L}{r_B}\Big)^{-m_B} \theta^{1-m_B}   \nonumber \\
      ~~~~~~~~~~~~~~~~~ + B_{z0}^2 \sigma_n^2 \Big(\frac{L}{r_n}\Big)^{-m_n} \theta^{1-m_n} \Big], \nonumber \\
       ~~~~~~~~~~~~~~~~~~~~~~~~~~~~~~~~~~~~~\frac{r_B}{L}<\theta<1,  \\
       2 \times 0.81^2 L^2 \Big [n_{e0}^2 \sigma_B^2 \Big(\frac{L}{r_B}\Big)^{-m_B}  \nonumber \\
       ~~~~~~~~~~~~~~~~~+ B_{z0}^2 \sigma_n^2 \Big(\frac{L}{r_n}\Big)^{-m_n} \Big],  \nonumber \\ 
       ~~~~~~~~~~~~~~~~~~~~~~~~~~~~~~~~~~~~~~ \theta>1>\frac{r_B}{L}. 
\end{subnumcases}
In the above expression, we consider the condition $r_n < r_B$. The opposite case with $r_B<r_n$ can be similarly formulated:
\begin{subnumcases}
     { D_\text{RM} (\theta)= \label{eq: msclex2}}
       2 \times 0.81^2 L^2 \Big [n_{e0}^2 \sigma_B^2  \Big(\frac{L}{r_B}\Big)^{m_B}\theta^{1+m_B}  \nonumber \\
       ~~~~~~~~~~~~~~~~~+ B_{z0}^2 \sigma_n^2  \Big(\frac{L}{r_n}\Big)^{m_n}\theta^{1+m_n} \Big ], \nonumber \\
        ~~~~~~~~~~~~~~~~~~~~~~~~~~~~\theta<\frac{r_B}{L}<\frac{r_n}{L}<1,    \\
       2 \times 0.81^2 L^2 \Big [n_{e0}^2 \sigma_B^2 \Big(\frac{L}{r_B}\Big)^{-m_B} \theta^{1-m_B}  \nonumber \\
       ~~~~~~~~~~~~~~~~~+ B_{z0}^2 \sigma_n^2  \Big(\frac{L}{r_n}\Big)^{m_n}\theta^{1+m_n} \Big],   \nonumber \\
        ~~~~~~~~~~~~~~~~~~~~~~~~~~~~\frac{r_B}{L}<\theta<\frac{r_n}{L}<1,    \\
       2 \times 0.81^2 L^2 \Big [n_{e0}^2 \sigma_B^2 \Big(\frac{L}{r_B}\Big)^{-m_B} \theta^{1-m_B}   \nonumber \\
      ~~~~~~~~~~~~~~~~~ + B_{z0}^2 \sigma_n^2 \Big(\frac{L}{r_n}\Big)^{-m_n} \theta^{1-m_n} \Big], \nonumber \\
       ~~~~~~~~~~~~~~~~~~~~~~~~~~~~~~~~~~~~~\frac{r_n}{L}<\theta<1,  \\
       2 \times 0.81^2 L^2 \Big [n_{e0}^2 \sigma_B^2 \Big(\frac{L}{r_B}\Big)^{-m_B}  \nonumber \\
       ~~~~~~~~~~~~~~~~~+ B_{z0}^2 \sigma_n^2 \Big(\frac{L}{r_n}\Big)^{-m_n} \Big],  \nonumber \\ 
       ~~~~~~~~~~~~~~~~~~~~~~~~~~~~~~~~~~~~~~ \theta>1>\frac{r_n}{L}. 
\end{subnumcases}
It reveals that the total $D_\text{RM}$ is the superposition of two components that are related to the fluctuations in electron density and the 
LOS component of magnetic field, respectively. 
In an exceptional situation with $r_n = r_B$ and $m_n = m_B$, namely, density and magnetic turbulence 
share the same spectral scaling law,
the expression of $D_\text{RM} (\theta)$ in Eq. \eqref{eq: comsfthe} can be recovered from either 
Eq. \eqref{eq: msclex} or Eq. \eqref{eq: msclex2}, and accordingly the CF parameters of RM density fluctuations can be more explicitly written as 
\begin{equation}
\begin{aligned}
  &  \sigma_\phi^2 = 0.81^2 (n_{e0}^2 \sigma_B^2 + B_{z0}^2 \sigma_n^2 ), \\
  &  r_\phi = r_n = r_B, \\
  &  m_\phi = m_n = m_B. 
\end{aligned}
\end{equation}
This is only valid under rather restrictive circumstances. 
More generally, one would expect that the spectra for fluctuating density and magnetic field are not aligned and 
the behavior of $D_\text{RM}$ depends on the relative importance between $B_{z0}^2 \sigma_n^2$ and $n_{e0}^2 \sigma_B^2$. 
For example, when
$B_{z0}^2 \sigma_n^2$ is significantly larger than $n_{e0}^2 \sigma_B^2$, that is, the relative density fluctuations are 
much stronger in comparison with relative magnetic field fluctuations, 
\begin{equation}\label{eq: relimp}
    \frac{\sigma_n^2}{n_{e0}^2} \gg  \frac{\sigma_B^2}{B_{z0}^2}, 
\end{equation}
density fluctuations dictate the behavior of $D_\text{RM}$. 
Both Eq. \eqref{eq: msclex} and \eqref{eq: msclex2} approximately go back to Eq. \eqref{eq: comsfthe}, and 
the CF parameters of RM density fluctuations are equivalent to 
\begin{equation}\label{eq: rmexp}
\begin{aligned}
  &  \sigma_\phi^2 = 0.81^2 B_{z0}^2 \sigma_n^2 , \\
  &  r_\phi = r_n, \\
  &  m_\phi = m_n. 
\end{aligned}
\end{equation}
In this situation,  magnetic field fluctuations are basically not responsible for the observed RM SFs.

As regards the SF of EM, combining Eq. \eqref{eq: sflp}, \eqref{eq: comsfthe}, and \eqref{eq: sepsfem} yields 
\begin{subnumcases}
     {  D_\text{EM}(\theta) = }
       8n_{e0}^2 \sigma_n^2 L^2 \Big(\frac{L}{r_n}\Big)^{m_n}\theta^{1+m_n}, \nonumber \\ 
       ~~~~~~~~~~~~~~~~~~~~~~~~~~~~~~~~\theta<\frac{r_n}{L}<1,\\
       8n_{e0}^2 \sigma_n^2 L^2 \Big(\frac{L}{r_n}\Big)^{-m_n} \theta^{1-m_n}, \nonumber \\ 
       ~~~~~~~~~~~~~~~~~~~~~~~~~~~~~~~~ \frac{r_n}{L}<\theta<1, \label{eq: demlang}\\
       8n_{e0}^2 \sigma_n^2 L^2 \Big(\frac{L}{r_n}\Big)^{-m_n}, \nonumber \\ 
       ~~~~~~~~~~~~~~~~~~~~~~~~~~~~~~~~ \theta>1>\frac{r_n}{L}.
\end{subnumcases}
The similarity between the form of the above $D_\text{EM}(\theta)$ and Eq. \eqref{eq: comsfthe} suggests that the EM density fluctuations
inherit the turbulence properties from the density field and their CF parameters are related as
\begin{equation}\label{eq: emdd}
\begin{aligned}
  &  \sigma_\phi^2 = 4n_{e0}^2  \sigma_n^2 , \\
  &  r_\phi = r_n, \\
  &  m_\phi = m_n. 
\end{aligned}
\end{equation}
If the situation described in Eq. \eqref{eq: relimp} is realized, we expect that
$ D_\text{RM}(\theta)$ and $D_\text{EM}(\theta)$ measured from the same region in the sky
exhibit a similar behavior in terms of the 
spectral slope and correlation scale corresponding to the break in slope. Their ratio 
\begin{equation}
    \frac{D_\text{RM} (\theta)}{D_\text{EM}(\theta) }  =\Big(\frac{0.81 B_{z0}}{2 n_{e0}}\Big)^2
\end{equation}
is associated with the mean plasma properties and determined by the ratio between the mean values of $B_z$ and $n_e$.

\section{Comparison with observations}\label{sec: comobs}

Observationally determined SFs of RMs allow a quantitative test of the above analysis. 
\citet{MS96} studied both RM and EM measurements for 38 extragalactic sources. 
The sample of source components are selected to have insignificant intrasource variations in RM and depolarization effects 
for the purpose of studying the Faraday rotation of the Galactic medium only. 
Another strong argument that the observed RM is dominated by the Galactic Faraday rotation is the evident dependence on the angular separation 
of the RM SF.
The measured RM SF $D_\text{RM}$ increases as a power-law-like function of angular scale and is characterized by a break in its spectrum 
around $0.1^\circ$. 
The least-squares fit to the data above $0.1^\circ$ gives 
\begin{equation} \label{eq: obsdrm}
    D_\text{RM} (\theta > 0.1^\circ) = (340 \pm 30) ~ [\theta(^\circ)]^{0.64\pm 0.06} ~\text{rad}^2 \text{m}^{-4}. 
\end{equation}
The three data points below $0.1^\circ$ are insufficient for achieving a reliable fitting, 
but they clearly indicate a more steepened spectrum.

These observed features of RM SFs appear to be consistent with the theoretical expectations presented in Section 2. 
To enable a quantitative comparison, we employ the parameters provided in 
\citet{MS96}. 
Since the observed Faraday rotation for their sample is dominated by the magnetized ISM of our Galaxy,
we adopt the average path length $2900$ pc through the ISM in the observed region as the depth $L$ of the Faraday screen.  
Besides,
the very few measurements at small angular separations do not allow an accurate determination of the break point in the SF spectrum. 
We then choose the ``outer scale'' of turbulence of $3.6$ pc suggested in
\citet{MS96}
as an experimental correlation scale $r_\phi$. 
The transition angle at the break is thus 
\begin{equation}
     \theta_\text{tr} = \frac{r_\phi} {L}= 0.0711^\circ. 
\end{equation}

\begin{figure}[htbp]
\centering
\includegraphics[width=9cm]{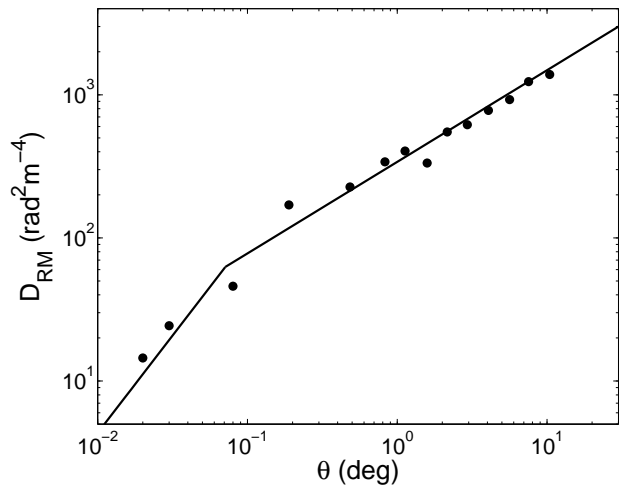}
\caption{ SFs of RMs vs. angular separation. The data points are taken from figure 5 in 
\citet{MS96}. The solid line corresponds to the theoretical formalisms Eq. \eqref{eq: comrmsa} and \eqref{eq: comrmla}. }
\label{figms96}
\end{figure}

Given the coincidence $\theta=R/L$ for this observation, and the determination of $L$ and $r_\phi$, 
we can compare Eq. \eqref{eq: obsdrm} with the functional form of $D_\text{RM}$ at $\theta > \theta_\text{tr}$ from Eq. \eqref{eq: comrmla},
which directly yields the CF index and variance of RM density fluctuations, 
\begin{equation}\label{eq: commsp}
    m_\phi = 1-0.64 = 0.36, ~~~
    \sigma_\phi = 5.5 \times 10^{-2} ~ \text{rad}~ \text{m}^{-2} \text{pc}^{-1}.
\end{equation}
Evidently, the measured $m_\phi$ does not coincide with the prediction by Kolmogorov turbulence, which entails
$m_\phi = 2/3$ instead (see Section \ref{sec: lp}). 
In fact, since the range $R>r_\phi$ (i.e., $\theta > \theta_\text{tr}$) corresponds to the inertial range of a shallow spectrum of RM density fluctuations, 
the resulting spectral index is (Eq. \eqref{eq: alaslw})
\begin{equation}\label{eq: msspesl}
\alpha = m_\phi -3 = 0.36-3 = -2.64. 
\end{equation}
Meanwhile, the identification of the inertial range over the scales larger than $r_\phi$ 
indicates that $r_\phi$ is actually the inner scale rather than the outer scale of turbulence. 
Below $r_\phi$, the damping effect efficiently suppresses the fluctuations 
and steepens the spectral tail in the dissipation range. 
If we assume that the same scaling as in the inertial range can still be used at scales below but in the vicinity of $r_\phi$,
inserting the values in Eq. \eqref{eq: commsp} into the expression of $D_\text{RM}$ at $\theta < \theta_\text{tr}$ 
from Eq. \eqref{eq: comrmsa} leads to 
\begin{equation}\label{eq: drmtos}
   D_\text{RM} (\theta < \theta_\text{tr})= 2.28\times10^3~ [\theta(^\circ)]^{1.36}  \text{rad}^2 ~\text{m}^{-4}.
\end{equation}
Fig. \ref{figms96} plots $D_\text{RM}$ from Eq. \eqref{eq: drmtos} at $\theta < \theta_\text{tr}$ and Eq. \eqref{eq: obsdrm} at 
$\theta > \theta_\text{tr}$, superposed with the observational data points taken from figure 5 in 
\citet{MS96}. 
It seems that both the amplitude and spectral slope of $D_\text{RM}$ given by Eq. \eqref{eq: drmtos} are in good agreement with the observational result. 
However, due to the limited number of close source pairs,
it is difficult to impose a strong constraint on the exact spectral slope in the 
dissipation range of turbulence. 
Nevertheless, the comparison illustrates that the theoretical model originally constructed by LP16
can satisfactorily interpret the observed RM SFs.

We exclude the possibility that $r_\phi$ is the outer scale of a steep spectrum, as in this case one would expect that the SF saturates at 
a constant value and flattens 
at $\theta > \theta_\text{tr}$. 
Otherwise, the observed SF spectrum over an extended range of angular scales beyond $\theta_\text{tr}$
severely challenges the model for the energy injection of turbulence.

From RM SFs alone, we are unable to identify the separate contributions due to density and magnetic field fluctuations. 
Hence, it is necessary to invoke SFs of EMs for an exclusive extraction of the electron density fluctuations.
We now turn to the observationally measured EM SFs. 
A power-law representation with a similar slope to that of RM SFs
fits the observations
\citep{MS96},
\begin{equation}\label{eq: obsem}
   D_\text{EM} (\theta) = (5.5 \pm 0.6) ~ [\theta(^\circ)]^{0.73\pm 0.08} \text{cm}^{-12} \text{pc}^2. 
\end{equation}
Due to the lower angular resolution of the EM data, the measurements of $D_\text{EM}$ at small angular scales are absent, 
which prevents the detection of the possible break in its spectrum and determination of the correlation scale of EM density fluctuations. 
But the fitting by Eq. \eqref{eq: obsem} is informative on the CF index of EM density fluctuations, which is also approximately the CF index of 
density fluctuations (see Eq. \eqref{eq: emdd}). 
Comparing Eq. \eqref{eq: comrmla} and \eqref{eq: obsem} gives the value 
\begin{equation}
 m_\phi = 1-0.73(\pm 0.08) = 0.27(\pm 0.08),
\end{equation}
close to that of RM density fluctuations (Eq. \eqref{eq: commsp}). 
The similarity between the slopes of RM and EM SFs strongly
suggests that the density fluctuations take a major part in composing the SF of RM. 
Presumably, the CF index and correlation scale obtained from RM SFs also match those quantities of density fluctuations
(see Eq. \eqref{eq: rmexp}), 
\begin{equation}\label{eq: presmr}
   m_n = 0.36, ~~~r_n = 3.6 ~\text{pc},
\end{equation}
and the EM SFs at $\theta>\theta_\text{tr}$ satisfy
\begin{equation}
   D_\text{EM} (\theta) = 5.5  ~ [\theta(^\circ)]^{0.64} \text{cm}^{-12} \text{pc}^2. 
\end{equation}
By comparing the above equation with Eq. \eqref{eq: demlang} and using $L= 2900$ pc and values in Eq. \eqref{eq: presmr},
we find
\begin{equation}
    n_{e0}\sigma_n = 3.47\times10^9 \text{m}^{-6}.
\end{equation}
Furthermore, under the condition of Eq. \eqref{eq: relimp}, 
we can safely neglect the terms associated with magnetic fluctuations in Eq. \eqref{eq: msclex} and \eqref{eq: msclex2}
due to their minor contribution. 
From Eq. \eqref{eq: rmexp} and \eqref{eq: commsp}, we get
\begin{equation}
   0.81 B_{z0} \sigma_n = \sigma_\phi  =  5.5 \times 10^{-2} ~ \text{rad}~ \text{m}^{-2} \text{pc}^{-1},
\end{equation}
indicative of the fact that the level of RM density fluctuations is determined by the joint strength of fluctuating 
electron density and mean magnetic field along the LOS.
Given the estimate of mean electron density $n_{e0}$, $\sigma_n$ and $B_{z0}$ can be both derived from the above two equations. 
However, the properties of the fluctuating component of magnetic field are poorly constrained by SFs of RMs. 

The above results suggest that the observed RM SFs 
originate from the underlying shallow spectrum of density fluctuations, with the break in the slope of the SF
corresponding to the inner scale of the density spectrum. 
Accordingly, the Faraday rotating medium that gives rise to the characteristics of RM SFs 
has an excess of dense structures at small scales comparable to $r_\phi$.

As a caveat to the applicability of the theoretical model, 
the uncertainties in the amplitude and correlation scale of turbulence obtained from observations 
are introduced by the choice of Faraday rotation depth $L$. 
The RM fluctuations traced by Galactic sources provide information about the small-scale turbulence within a specific region in 
the Galaxy. 
The Faraday rotating medium extends from the observer to sources and thus the sources' distances can be used 
as the estimate of $L$. 
Observations of extragalactic sources with marginal intrinsic Faraday rotation as in 
\citet{MS96}
can bring forth large-scale features of ISM turbulence. 
The depth of the Faraday screen is the path length throughout the Galaxy, and the sources' distances are irrelevant.
But in the case with dominant intrinsic Faraday rotation, 
provided that the multiple RM components within one source can be resolved, 
the properties of turbulence in the source region are probed. 
Apart from $L$, the source's distance is also involved in the analysis
(see Eq. \eqref{eq: extrintr}).

\section{Conclusions and discussion}

Following the model SF of RM fluctuations put forth by 
LP16,
we proceed to carry out the SF analysis of RM and EM by separating the contributions from 
fluctuations of electron density and the LOS component of magnetic field, and we 
assess their relative importance in determining the form of SFs of RMs as a function of angular scales. 
We found that the SF of RM can be considered as a sum of two SFs stemming from the fluctuations of electron density and magnetic field, respectively. 
The turbulent spectrum of the density fluctuations can be extracted from the SF measurement of EM 
provided that the angular resolution is sufficient.
Applying the analysis to observationally determined SFs of RMs shows that 
the model SF can consistently interpret the shape of SFs on different ranges of angular scales. 
Similar behavior of RM and EM SFs suggests that the observed SF of RM is mainly guided by fluctuating electron density and thus
incapable of probing the nature of turbulent magnetic field.

As an observational fact,
the changing shape of SFs of RMs with angular scales was earlier interpreted as arising from a transition from three-dimensional to 
two-dimensional turbulence 
\citep{MS96}, 
or two spatially separate Faraday screens 
\citep{Hav04}.
Without the intervention of two turbulent structures or two Faraday rotation regions, 
the power-law CF for RM density 
fluctuations introduced in 
LP16 can naturally reproduce the observed features and explain the broken power-law shape of the RM SF.

In earlier works, magnetic field is assumed to be frozen in matter, and thus both density and magnetic field fluctuations 
adhere to an identical power-law spectrum with the same spectral index and inner and outer scales. 
But this conjecture has been definitely rejected by observations on RM fluctuations.
The Kolmogorov spectrum appears to be too steep to account for the shallow slope especially at scales larger than the 
spectral break,  and it alone cannot serve as a satisfactory turbulent model to fit the 
various slopes of SFs from different observations. 
In fact, it has been known that the conventional flux-freezing concept breaks down in realistic MHD turbulence as 
the diffusion of magnetic field lines is mediated by 
fast magnetic reconnection, which is an intrinsic process inherent in MHD turbulence 
\citep{Laz05, Laz11,LEC12, EyNa13}.
It is more plausible that density and magnetic field fluctuations conform to distinct power spectra of turbulence. 
The dominant one between them is more important in determining the shape of the resultant SF spectrum of RM.

The similarity between the behavior of RM and EM SFs revealed by observations 
(e.g. \citealt{MS96, Hav04})
indicates that the major contribution to the measured RM SFs comes from electron density fluctuations, 
which tend to follow a shallow spectrum of turbulence down to the dissipation scale corresponding to the spectral break of 
the RM SF.
Both theoretical considerations and numerical simulations suggest that a shallow spectrum of density field can arise in 
compressible turbulent flows.
Compressibility leads to the formation of clumpy density structures, with condensations embedded in 
relatively diffuse regions
\citep{BLC05, Kri07, KL07,Fal14}. 
The coupling between this density structure and local turbulent motions results in a steeper 
velocity power spectrum with a slope of $\sim -2$, but a much shallower power spectrum of the density field than
the Kolmogorov $-5/3$ scaling for one-dimensional spectra. 
Moreover,
the shallowness of the spectral slope of density fluctuations is strongly affected by magnetic field strength in subsonic turbulence 
and by sonic Mach number $M_s$ in supersonic turbulence
\citep{KL07}. 
For example, 
\citet{KL07}
observed a slope of $\sim -0.5$ for the one-dimensional density spectrum obtained from the simulated 
supersonic turbulence with $M_s = 7$, which is even shallower than the spectral slope indicated from the 
RM SFs measured by
\citet{MS96} (see Eq. \eqref{eq: msspesl}). 
The ISM is highly inhomogeneous with dense structures accumulated by shocks on small scales. 
Depending on the local compressibility,
the density fluctuations and the resultant RM SFs can have spatially diverse spectral slopes.

In contrast to density, magnetic field is better coupled with turbulent velocity field and regulates the turbulence
properties, e.g., scale-dependent turbulent anisotropy, as the turbulent energy cascades down from large to small scales. 
The ``Big Power Law in the Sky" extending from $10^{17}$ m to $10^6$ m 
\citep{Armstrong95,CL10}
suggests that the interstellar turbulence has a Galactic-scale ($>100$ pc) energy injection source and cascades toward very small scales. 
At the parsec scale of this global turbulent cascade, magnetic field fluctuations can have a lower
amplitude compared with the enhanced amplitude of density fluctuations. These excessive density fluctuations are 
thus manifest by dominating the composite RM density fluctuations. 
The candidate regions corresponding to the localized enhanced turbulence in density field can be 
the extended envelopes of H {\small \uppercase\expandafter{\romannumeral2}} regions and the warm ionized medium of the 
\citet{MO77}
model in the Galactic plane 
\citep{Span91, Armstrong95}.


\citet{CL10} carried out a remarkable extension of the Big Power Law in the Sky in 
\citet{Armstrong95}
up to $10^{17}$ m by using the data of the Wisconsin H$\alpha$ Mapper (WHAM)
and demonstrate a universal spectrum of interstellar density fluctuations consistent with Kolmogorov turbulence. 
On the contrary, observations of RM fluctuations at low Galactic latitudes
(e.g. \citealt{Sim86, Cle92, Hav04,Hav08})
imply a shallower spectrum than the Kolmogorov one.
As a possible understanding of this contradiction,  
for LOSs that traverse the supersonic turbulent flows through the Galactic plane, 
the RM results are mainly governed by density fluctuations, which we do not expect to be compatible with the Kolmogorov spectrum 
as they are more independent from turbulent velocity and sensitive to the local compressibility of the medium. 
\citet{CL10} analyzed the WHAM data at high Galactic latitudes and 
avoided the contamination from the H {\small \uppercase\expandafter{\romannumeral2}} regions in the Galactic plane.
As a result, they achieved one universal turbulent spectrum throughout the Galaxy. 
Their result is also in agreement with the study of velocity turbulence by using the velocity coordinate spectrum technique in 
\citet{CLS10}.
So great caution is needed when one uses RM fluctuations to probe the turbulent magnetic fields in the Galaxy, not only because 
the fluctuating density instead of fluctuating magnetic field can dominate the observed RM SF behavior, 
but also because more complexity can be introduced by additional structures of turbulence embedded in the observed region. 
They both hinder the recovery of the underlying spectrum of magnetized turbulence.

We identify the transition scale on the order of $1$ pc at the spectral break of RM SFs as the inner scale of 
a shallow electron density spectrum over larger scales, instead of the injection scale of RM fluctuations at smaller scales as determined by 
\citet{Hav04,Hav08}. 
In a partially ionized ISM, the cascade of MHD turbulence is severely damped due to ion-neutral collisions. 
Below the scale where neutral fluid decouples from ion-electron fluid, 
the MHD turbulence in ion-electron fluid is efficiently suppressed by ion-neutral collisional damping, but the cascade of hydrodynamic turbulence in neutrals proceeds 
down to the viscous cutoff. 
The dissipation scale inferred from RM SFs we obtained in this work is consistent with the ion-neutral collisional damping scale of MHD turbulence 
in the warm neutral phase of the ISM calculated by
\citet{Xu15}.
Plausibly, this consistency could suggest that the inner scale to electron density fluctuations of $\sim 1$ pc characterizes the typical and also minimum 
scale of discrete structures of excess electron densities in the observed region, 
while smaller condensations mostly consist of neutrals and are driven by gravitational contraction instead of supersonic motions. 
\\
\\
The authors thank the anonymous referee for the valuable comments.
This work is partially supported by the National Basic Research Program (973 Program) of China under grant No. 2014CB845800.

\appendix

\section{Calculations for the CFs of RM and EM densities}
\label{sec:app}

The CFs of RM and EM densities are calculated as 
\begin{equation}
\begin{aligned}
 & \xi_\phi(\text{RM})  \\
 = & 0.81^2 \langle (n_e B_z) (\bm{X_1}, z_1)  ~ (n_e B_z) (\bm{X_2}, z_2) \rangle \\
 = & 0.81^2 \Big [ n_{e0}^2 B_{z0}^2  
     + n_{e0}^2  \langle \delta B_z(\bm{X_1}, z_1) \delta B_z(\bm{X_2}, z_2) \rangle 
      + B_{z0}^2 \langle \delta n_e(\bm{X_1}, z_1) \delta n_e(\bm{X_2}, z_2) \rangle \\
   &  + \langle \delta n_e(\bm{X_1}, z_1) \delta B_z(\bm{X_1}, z_1) \delta n_e(\bm{X_2}, z_2) \delta B_z(\bm{X_2}, z_2)   \rangle \Big] \\
 =& 0.81^2 (n_{e0}^2 B_{z0}^2 + n_{e0}^2  \widetilde{\xi}_B +  B_{z0}^2  \widetilde{\xi}_n +  \widetilde{\xi}_{nB} ) \\
 =& 0.81^2 (n_{e0}^2 B_{z0}^2 + n_{e0}^2  \widetilde{\xi}_B +  B_{z0}^2  \widetilde{\xi}_n +  \widetilde{\xi}_{n}  \widetilde{\xi}_B), 
\end{aligned}
\end{equation}
and 
\begin{equation}
\begin{aligned}
 & \xi_\phi(\text{EM})   \\
 =& \langle (n_e^2) (\bm{X_1}, z_1)  ~ (n_e^2) (\bm{X_2}, z_2) \rangle \\
 =& n_{e0}^4 +  2 n_{e0}^2 \langle \delta n_e  ^2 \rangle 
  + \langle \delta n_e (\bm{X_1}, z_1) ^2 \delta n_e (\bm{X_2}, z_2) ^2\rangle 
  + 4 n_{e0}^2  \langle \delta n_e (\bm{X_1}, z_1)\delta n_e (\bm{X_2}, z_2) \rangle \\
 =& n_{e0}^4 +  2 n_{e0}^2 \langle \delta n_e  ^2 \rangle  + \widetilde{\xi}_{nn} + 4n_{e0}^2 \widetilde{\xi}_n \\
 =& (n_{e0}^2 + \langle \delta n_e^2   \rangle)^2 + 2  \widetilde{\xi}_n^2 + 4n_{e0}^2  \widetilde{\xi}_n .
\end{aligned}
\end{equation}
In our derivations above, 
the three-point correlations are neglected under the assumption that the turbulent fluctuations are Gaussian processes,
and the turbulent density and magnetic field are assumed to be uncorrelated
\citep{MS96}. 
Following the four-point correlations as employed in 
\citet{MS96}, the expressions of $\widetilde{\xi}_{nB}$ and $\widetilde{\xi}_{nn}$ in the above equations are 
\begin{equation}
  \widetilde{\xi}_{nB}=  \widetilde{\xi}_{n}  \widetilde{\xi}_B, ~~
   \widetilde{\xi}_{nn} = \langle \delta n_e^2   \rangle^2 + 2 \widetilde{\xi}_n^2.
\end{equation}

\bibliographystyle{apj.bst}
\bibliography{yan}

\end{document}